\documentclass[sigconf]{acmart}

\AtBeginDocument{%
  \providecommand\BibTeX{{%
    \normalfont B\kern-0.5em{\scshape i\kern-0.25em b}\kern-0.8em\TeX}}}

\usepackage{svg}
\usepackage{balance}
\usepackage[breakable]{tcolorbox}
\usepackage{longtable}
\usepackage{bbding}
\usepackage{multirow}
\usepackage{balance} 
\usepackage{etoolbox} 
\usepackage{amsmath}
\usepackage{hyperref}
\usepackage{colortbl}
\usepackage{enumitem}
\usepackage{pifont}
\usepackage{float}    
\usepackage{subfig}   
\usepackage{overpic}  
\usepackage{xspace}
\usepackage{cleveref}

\newcommand{\ourtool}{\text{PatchFinder}\xspace}

\newcommand{\numpatch}{{533}\xspace}
\newcommand{\numconfirm}{{482}\xspace}
\newcommand{\mrr}{\text{0.7951}\xspace}

\newcommand{\lstbg}[3][0pt]{{\fboxsep#1\colorbox{#2}{\strut #3}}}

\usepackage{array}  
\newcolumntype{B}{>{\color{black}}c}

\newcommand{\revised}[1]{\textcolor{black}{#1}}

\usepackage{listings}
\crefname{lstlisting}{listing}{listings}
\Crefname{lstlisting}{Listing}{Listings}
\usepackage{xcolor}

\definecolor{codegreen}{rgb}{0,0.6,0}
\definecolor{codegray}{rgb}{0.5,0.5,0.5}
\definecolor{codepurple}{rgb}{0.58,0,0.82}
\definecolor{shallowred}{rgb}{1,0.8,0.8}
\definecolor{verylightgray}{rgb}{0.97, 0.97,0.97}

\lstdefinestyle{mystyle}{
    language=C++, 
    commentstyle=\color{codegreen},
    keywordstyle=\color{blue},
    stringstyle=\color{codepurple},
    basicstyle=\ttfamily\scriptsize\bfseries,
    breakatwhitespace=false,
    breaklines=true,              
    captionpos=b,                    
    keepspaces=true,                 
    numbers=left,                    
    numbersep=5pt,                  
    showspaces=false,                
    showstringspaces=false,
    frame=single,                    
    xleftmargin=0.05\linewidth,
    xrightmargin=0.05\linewidth,
    framexleftmargin=0.05mm,            
    framexrightmargin=0.05mm,           
    framextopmargin=0mm,             
    framexbottommargin=0mm,    
    showtabs=false,                  
    tabsize=2,
    escapeinside={(*@}{@*)},
}
\lstdefinelanguage{diff}{
	frame=single,
    breakatwhitespace=false,
    breaklines=true,
	basicstyle=\ttfamily\scriptsize, 
	morecomment=[f][\color{red}]{---}, 
	morecomment=[f][\color{codegreen}]{+++},
	morecomment=[f][\lstbg{red!20}]{-},
	morecomment=[f][\lstbg{green!20}]{+},
	morecomment=[f][\color{blue}]{@@},
    xleftmargin=0.03\linewidth,
    xrightmargin=0.005\linewidth,
    framexleftmargin=0.05mm,            
    framexrightmargin=0.0mm,           
    framextopmargin=0mm,             
    framexbottommargin=0mm,
    numbers=left,                    
}

\setcopyright{acmcopyright}
\copyrightyear{2018}
\acmYear{2018}
\acmDOI{XXXXXXX.XXXXXXX}

\acmConference[ISSTA 2024]{ACM SIGSOFT International Symposium on Software Testing and Analysis}{16-20 September, 2024}{Vienna, Austria}
\copyrightyear{2024}
\acmYear{2024}
\setcopyright{rightsretained}
\acmConference[ISSTA '24]{Proceedings of the 33rd ACM SIGSOFT International
Symposium on Software Testing and Analysis}{September 16--20, 2024}{Vienna,
Austria}
\acmBooktitle{Proceedings of the 33rd ACM SIGSOFT International Symposium
on Software Testing and Analysis (ISSTA '24), September 16--20, 2024, Vienna,
Austria}\acmDOI{10.1145/3650212.3680305}
\acmISBN{979-8-4007-0612-7/24/09}

\begin{document}

\title{\ourtool: A Two-Phase Approach to Security Patch Tracing for Disclosed Vulnerabilities in Open-Source Software}


\author{Kaixuan Li}
\affiliation{%
  \institution{East China Normal University}
  \city{Shanghai}
  \country{China}
}
\affiliation{
  \institution{Continental-NTU Corporate Lab, Nanyang Technological University}
  \city{Singapore}
  \country{Singapore}
}
\email{kaixuan.li@ntu.edu.sg} 

\author{Jian Zhang}
\authornote{Corresponding author.}
\affiliation{%
  \institution{Nanyang Technological University}
    \city{Singapore}
  \country{Singapore}
}
\email{jian_zhang@ntu.edu.sg}

\author{Sen Chen}
\affiliation{
  \institution{College of Intelligence and Computing, Tianjin University}
  \city{Tianjin}
  \country{China}
}
\email{senchen@tju.edu.cn}

\author{Han Liu}
\affiliation{%
  \institution{Shanghai Key Laboratory of Trustworthy Computing, East China Normal University}
  \city{Shanghai}
  \country{China}
}
\email{hanliu@stu.ecnu.edu.cn}

\author{Yang Liu}
\affiliation{%
  \institution{Nanyang Technological University}
    \city{Singapore}
  \country{Singapore}
}
\email{yangliu@ntu.edu.sg}

\author{Yixiang Chen}
\affiliation{%
  \institution{Shanghai Key Laboratory of Trustworthy Computing, East China Normal University}
  \city{Shanghai}
  \country{China}
}
\email{yxchen@sei.ecnu.edu.cn}



\begin{abstract}
Open-source software (OSS) vulnerabilities are increasingly prevalent, emphasizing the importance of security patches. However, in widely used security platforms like NVD, a substantial number of CVE records still lack trace links to patches. Although rank-based approaches have been proposed for security patch tracing, they heavily rely on handcrafted features in a single-step framework, which limits their effectiveness.

In this paper, we propose \ourtool, a two-phase framework with end-to-end correlation learning for better-tracing security patches. In the \textbf{initial retrieval phase}, we employ a hybrid patch retriever to account for both lexical and semantic matching based on the code changes and the description of a CVE, to narrow down the search space by extracting those commits as candidates that are similar to the CVE descriptions. 
Afterwards, in the \textbf{re-ranking phase}, we design an end-to-end architecture under the supervised fine-tuning paradigm for learning the semantic correlations between CVE descriptions and commits. In this way, we can automatically rank the candidates based on their correlation scores while maintaining low computation overhead. 
We evaluated our system against 4,789 CVEs from 532 OSS projects. The results are highly promising: \ourtool achieves a Recall@10 of 80.63\% and a Mean Reciprocal Rank (MRR) of 0.7951. Moreover, the Manual Effort@10 required is curtailed to 2.77, marking a 1.94 times improvement over current leading methods. 
\revised{When applying \ourtool in practice, we initially identified \numpatch patch commits and submitted them to the official, \numconfirm of which have been confirmed by CVE Numbering Authorities.}
\end{abstract}
\maketitle

\section{Introduction}
Open-source software (OSS) is fundamental to the industrial applications and software community. 
This widespread adoption, however, comes with security challenges. The open and accessible nature of OSS has inadvertently led to a surge in security vulnerabilities~\cite{zhang2023mitigating,zhao2023software,liu2022demystifying}. The notorious vulnerabilities such as Log4Shell~\cite{log4shell} and Spring4Shell~\cite{spring4shell} have put millions at risk of data theft and service denials, reducing trust in the OSS ecosystem. 

To facilitate understanding and remediation for vulnerabilities, public security platforms,  including the Common Vulnerabilities and Exposures (CVE) and the National Vulnerability Database (NVD), provide details (e.g., descriptions~\cite{guo2022detecting}) on disclosed software vulnerabilities and links to relevant patches for mitigation. However, a recent study revealed that almost 57\% of CVEs lack trace links to patches, and only 12\% of commits in OSS reference the corresponding CVE-IDs~\cite{Tan-ccs21}. One fact is that the maintainers (e.g.,  CVE Numbering Authorities, CNAs) may not update the trace link on CVE/NVD even though the vulnerability has been fixed. This emphasizes the urgent necessity of security patch tracing to enhance the platforms' utility for developers and users.

Yet, pinpointing the exact security patches remains a significant challenge. These patches, which are commits made by OSS developers, are sparsely dispersed throughout individual code repositories. For instance, as of September 2023, the Linux kernel has over 1,215,313 commits, but typically only one of those commits is the patch for a vulnerability. Previous efforts have attempted to search them by leveraging auxiliary information from vulnerability databases, like CVE-IDs in commits~\cite{perl2015vccfinder,kim2017vuddy} and external reference URLs from CVE/NVD pages~\cite{Xu-fse22}. However, as evident from the aforementioned statistics, these methods often fall short due to their matching-based nature.

In response to this challenge, the research community has shifted to rank-based methods to pair vulnerabilities with patches. PatchScout~\cite{Tan-ccs21} and VCMatch~\cite{wang2022vcmatch} are two representative works that build on pair-wise ranking and point-wise ranking respectively. The common practice involves manually defining and extracting features from descriptions and commits, for example, the number of shared words between the description and commit message. Subsequently, the resulting feature vector is used to train a ranking model or classifier. 
While these rank-based methods offer advantages over traditional matching-based techniques, they also have a set of limitations. 
First, these methods \textbf{\textit{predominantly depend on handcrafted features}} and neglect the rich semantic information. 
\revised{
Concretely, \ding{172} existing models like PatchScout and VCMatch rely heavily on direct lexical matches (20/22 and 34/36, respectively). 
The predominantly word-based similarities fall short of capturing semantics including semantic-equivalent phrases and implicit patch information from both sides (CVE descriptions \& Commits). 
\ding{173} While two unsupervised features used in existing works provide some benefits, they are not task-specific and cannot fully capture the varying correlations of semantics that are crucial for the task. This limitation can compromise their effectiveness and generality. 
}
Second, due to \textbf{\textit{the vast gap between the number of developmental commits and security patches}}, existing approaches cannot be easily adapted to learn \revised{supervised features. 
The current single-step framework, which involves either directly ranking or classifying commits, presents challenges in effectively training a supervised learning model for the identification of potential patches. }

In this paper, we introduce a two-phase framework including initial retrieval and re-ranking, which enables us to learn \revised{supervised semantics between commits and CVE descriptions}  in an end-to-end manner. Our approach, namely \ourtool, leverages both strengths of Information Retrieval (IR) and Large Language Models (LLMs) to capture lexical and semantic information and commit-related domain knowledge from the two phases respectively. 
In the initial retrieval phase, we employ a hybrid commit retriever for narrowing down the search space, which consists of TF-IDF (Term Frequency-Inverse Document Frequency)~\cite{chowdhury2010introduction} and a pre-trained CodeReviewer. That is, we extract commit candidates that are lexically and semantically similar to the CVE descriptions. 
In this way, we can sharply narrow the gap as previously discussed. Certainly, this algorithm cannot perfectly understand the correlations. Nevertheless, the refined dataset allows us to mitigate it via supervised learning, which fundamentally differs from existing approaches. 
In the re-ranking phase, inspired by the success of LLMs on Natural Language Processing (NLP) tasks, we take the CodeReviewer as the foundation model that is tailored specifically for understanding code changes. To unlock its potential, we design an end-to-end architecture under the fine-tuning paradigm for learning the semantic correlations between CVE descriptions and commits. Specifically, we simultaneously encode the description and one commit (including its message and code diff) to obtain two vector representations from the final layer of the CodeReviewer encoder. The vectors are subsequently concatenated into a single vector, which is then fed into a linear classifier to determine whether they are related or not. Together, the hybrid candidate retrieval using TF-IDF and pre-trained CodeReviewer, combined with the semantic correlation captured by fine-tuned CodeReviewer, provides a robust solution to the challenge of tracing security patches.

To comprehensively evaluate our approach, we enrich existing datasets by incorporating newly released CVEs from CVE/NVD, enlarging the dataset from \revised{1,669 to 4,789 CVE entries} (4870 patch commits). We then perform extensive experiments on this expanded dataset.
The experimental results show that \ourtool is highly effective and significantly outperforms all baselines. Specifically, \ourtool boasts a Recall@10 of $80.63\%$ and a Mean Reciprocal Rank (MRR) of \mrr. Moreover, the Manual Efforts@10 in real-world scenarios are curtailed to just $2.77$, marking a notable improvement over the state-of-the-art (SOTA) by 1.94 times. 
Notably, when deployed in real-world projects, \revised{\ourtool successfully identified \numpatch new security patches with an average rank of $1.65$. Of these, \numconfirm has been confirmed by CNAs. }

Our main contributions are as follows.
\begin{itemize}[leftmargin=10pt]
    \item We present \revised{a two-phase framework} for security patch locating: a hybrid initial retrieval phase to refine the search space, followed by a re-ranking phase to learn the correlations between CVE descriptions and patches.
    \item We design \ourtool, a comprehensive system that combines the strengths of TF-IDF and CodeReviewer to effectively retrieve potential patch commits while capturing supervised semantics from both descriptions and commits. 
    \item Through extensive evaluations against $4,789$ CVEs from $532$ OSS projects, we demonstrate the effectiveness of \ourtool with a Recall@10 of $80.63\%$ and an MRR of $\mrr$, outperforming state-of-the-art methods and reducing manual efforts significantly. Additionally, with the help of \ourtool, we found and submitted \numpatch new security patches for the CVE official, of which \numconfirm ones have been confirmed and updated by the CNAs.
    \item We have released all of our code and data on our website~\cite{website} for reproduction and further research.
\end{itemize}


\section{Background and Motivation}\label{sec:background}

\subsection{Large Language Models (LLMs)}

Pre-trained language models like BERT~\cite{devlin2018bert}, GPT~\cite{radford2018improving}, \revised{Llama2}~\cite{touvron2023llama}, and T5~\cite{t5} have significantly advanced NLP tasks. 
These models adopt a pre-training and then fine-tuning paradigm to develop transferable language representations. 
This paradigm has been adapted to programming languages with models such as CodeBERT~\cite{feng2020codebert}, \revised{CodeLlama~\cite{roziere2023code}}, CodeT5~\cite{wang2021codet5}, and CodeReviewer~\cite{codereview-fse22}. These models have demonstrated remarkable effectiveness, achieving SOTA performance on various code-related tasks and significantly improving code understanding and generation capabilities.

In specific, CodeReviewer~\cite{codereview-fse22} is a pre-trained Transformer-based encoder-decoder language model based on CodeT5~\cite{wang2021codet5}. It was pre-trained with code change and code review data collected from OSS projects on GitHub to support code review tasks for the nine most popular programming languages.  Compared with other LLMs, CodeReveiwer has the following characteristics: 
\textit{1) Purpose-built for Code Change Analysis:} Unlike general-purpose models or those optimized for a broader range of tasks, CodeReviewer is specifically tailored for analyzing code changes. This makes it an apt choice for understanding the semantics of commits, which is pivotal for our task.
\textit{2) Pre-training on CodeT5:} CodeReviewer's foundation on CodeT5 means it has benefited from vast amounts of code data during its pre-training phase. This gives it a knowledge advantage over other models that might not have had access to similar training data or might not be as recent as CodeT5.



\subsection{Problem Definition}
Since it is labor-intensive to trace the security patches for disclosed software vulnerabilities of CVEs, our objective is to design a model that can automatically identify the patches from OSS projects. We view the process as a ranking problem that ranks the commits in OSS projects based on their correlations with reported CVEs. Ideally, when provided with a CVE, the model should rank the associated security patch as high as possible. The model could take the CVE description, commit messages, and code changes (diffs) as input. It then generates a ranked list of commits for the given CVE, indicating the likelihood of each commit being the relevant patch. 

\noindent\textbf{Input Data:}
\textit{CVE Descriptions} and \textit{Commits}. Let \( \mathcal{D} \) be the set of CVE descriptions, such that \( \mathcal{D} = \{d_1, d_2, \ldots, d_{|\mathcal{D}|}\}\). For each description \( d_i \) in \( \mathcal{D} \), we have a corresponding set of commits \( \mathcal{C}_i = \{c_1, c_2, \ldots, c_{|\mathcal{C}_i|}\}\). Each commit \( c_j \) in \( \mathcal{C}_i \) is represented as a tuple containing its message and code diff, i.e., \( c_j = (msg_j, diff_j) \). 

\noindent\textbf{Output Data:}
\textit{Ranked Commits.} For a given CVE description \( d_i \) and its associated commits in \( \mathcal{C}_i \), the model produces a ranking vector \( \mathbf{R}_i = [r_1, r_2, \ldots, r_{|\mathcal{C}_i|}]\). This vector indicates the likelihood of each commit being the patch for \( d_i \), allowing the commits to be sorted based on their rankings.




\subsection{Motivating Example}

\begin{lstlisting}[language=diff,caption=A motivating example for CVE-2015-1867.,label={lst:motivate_eg}, float=t]
CVE description of CVE-2015-1867:
Pacemaker before 1.1.13 does not properly evaluate added nodes, which allows remote read-only users to gain privileges via an acl command.
****************************************************************
Commit message:
    Fix: acl: Do not delay evaluation of added nodes in some situations
    It is not appropriate when the node has no children as it is not a placeholder
    
Code diff:
diff --git a/lib/common/xml.c b/lib/common/xml.c
index f3dd35b7a..716f053f8 100644
--- a/lib/common/xml.c
+++ b/lib/common/xml.c
@@ -1020,13 +1020,16 @@ __xml_acl_post_process(xmlNode * xml)
+        char *path = xml_get_path(xml);
-        /* Always allow new scaffolding, ie. node with no 
-           attributes or only an 'id' */
+        /* Always allow new scaffolding, ie. node with no 
+           attributes or only an 'id' Except in the ACLs section
+         */
-            if (strcmp(prop_name, XML_ATTR_ID) == 0) {
+            if (strcmp(prop_name, XML_ATTR_ID) == 0 && 
+                strstr(path, "/"XML_CIB_TAG_ACLS"/") == NULL) {
@@ -1035,7 +1038,6 @@ __xml_acl_post_process(xmlNode * xml)
-                char *path = xml_get_path(xml);
@@ -1046,6 +1048,7 @@ __xml_acl_post_process(xmlNode * xml)
+        free(path);
\end{lstlisting}

The example in~\Cref{lst:motivate_eg} illustrates the challenge of associating a CVE description with its corresponding patch commit~\cite{pacemaker}. The description for CVE-2015-1867~\cite{CVE-2015-1867} (Line 2) hints at a vulnerability in \textit{pacemaker} for versions below \texttt{1.1.13}, but lacks specifics such as its exact location including the function name or file name. Moreover, the commit message (Lines 5-6) provides a hint about the root cause but expresses it differently and does not mention the exploitation of the vulnerability.
Existing SOTA tools like PatchScout and VCMatch predominantly rely on commit messages, predefined vulnerability type mappings, and handcrafted features~\cite{Tan-ccs21,wang2022vcmatch}. Such an approach can be limiting, especially when faced with ambiguous CVE descriptions that do not directly match commit messages. Solely relying on token-based features without considering the semantic nuances present in the message and diff can lead to inaccuracies.

In contrast, a more in-depth analysis of the commit, as demonstrated in our motivating example, reveals critical semantic insights that are essential for accurate tracing. Specifically, in Line 14 and Line 26, there is an addition of a new variable \texttt{path} which retrieves the path of the XML node. This suggests that the location or context of the XML node in the document might be significant for the fix. It then modifies a comment to specify an exception for the ACLs section (Lines 15-19), indicating that the behavior of allowing new scaffolding nodes is being refined. The condition in Lines 20-22 is enhanced to include a check that ensures the current XML path is not within the ACLs section, as indicated by the string \texttt{XML\_CIB\_TAG\_ACLS}. This is a direct response to the vulnerability mentioned in the CVE description, which \textit{allows privileges escalation via an {acl} command}. \revised{The removal of the \texttt{path} variable assignment is shown at Line 24}, which is now redundant due to its declaration and assignment at the beginning of the block. 

It is evident that while the commit message primarily reflects the CVE description, the semantic information in the code diff provides a clearer picture of how the vulnerability is addressed. This underscores the importance of analyzing complex semantics from code diffs in conjunction with commit messages to accurately link CVE descriptions with their corresponding patches.

\section{Approach}\label{sec:approach}

\begin{figure*}
    \centering
    \includegraphics[width=0.8\textwidth]{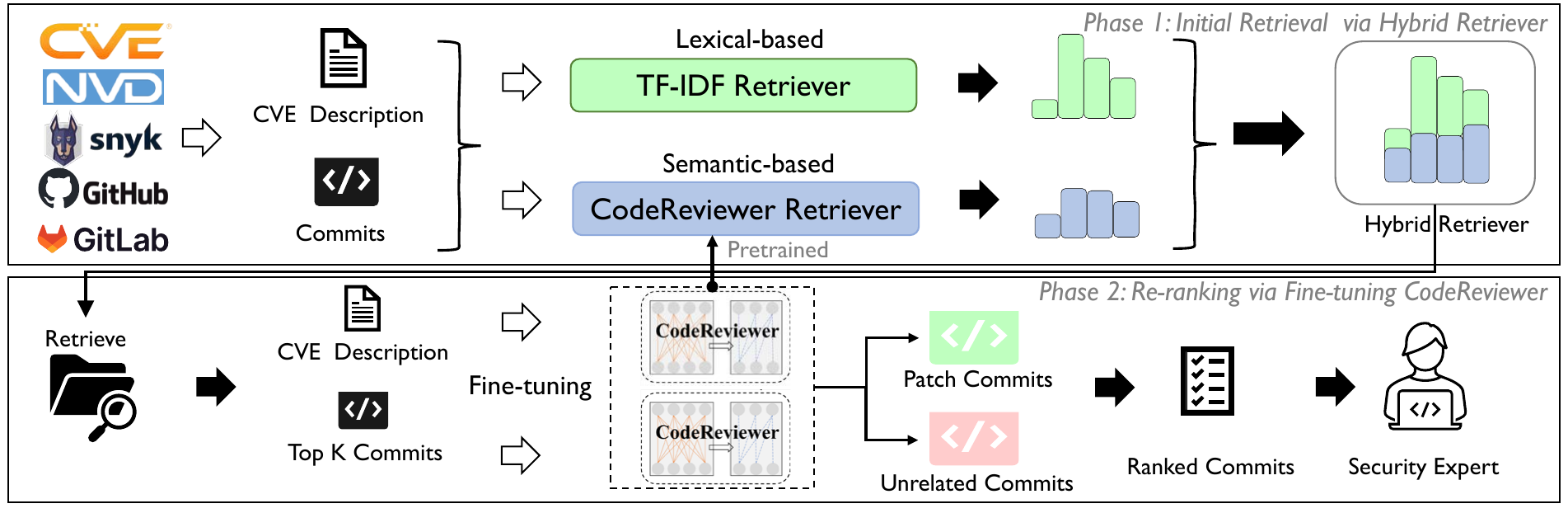}
    \caption{Overview of our approach.}
    \label{fig:overview}
\end{figure*}

\subsection{Overview}
In this section, we present an overview of our approach designed to trace security patches for disclosed vulnerabilities \revised{from the NVD/CVE websites}. 
As highlighted in earlier discussions, the primary challenge lies in the implicit correlations between CVE descriptions and commits, which necessarily require understanding the semantics of both sides. 
Theoretically, we recognized the challenges posed by direct retrieval using a fine-tuned LLM in extensive while extremely imbalanced datasets could diffuse the model's attention, reducing its effectiveness in accurately identifying relevant patches. 
To minimize the training loss, the features of the minority class (i.e. security patches) are easily treated as noise and are often ignored. Thus, there is a high probability of misclassification of the minority class as compared to the majority class (i.e. non-patch commits). 
Modifying the loss function alone results in a substantial computational burden during the fine-tuning process of LLMs on datasets exceeding 20 million entries. 



To tackle these drawbacks, as illustrated in Figure \ref{fig:overview}, we propose a novel two-phase framework called \ourtool, comprising initial retrieval and re-ranking. 
Initially, we retrieve the top candidates of patches from the commits based on lexical and semantic information, which helps eliminate a majority of unrelated commits, such as developmental code changes. On top of that, we design an end-to-end architecture based on LLMs to effectively capture the precise semantics of CVE descriptions and commits. 
Technically, in the \textit{initial retrieval} phase, we commence by preprocessing the CVE descriptions, commit messages, and code diffs. Subsequently, we employ the TF-IDF and pre-trained CodeReviewer to compute similarity scores between the given CVE description and each code commit at lexical and semantic levels, respectively. 
Note that this phase is not merely a preliminary step for narrowing down the search space but is critical for ensuring that the re-ranking phase can operate with enhanced focus and accuracy (further details in~\Cref{sec:tfidf}). 
Transitioning to the re-ranking phase, we harness the capabilities of LLMs, specifically CodeReviewer, which was pre-trained for code change analysis and defect understanding~\cite{codereview-fse22}.
By fine-tuning CodeReviewer with the top $k$ commits retrieved in the initial retrieval phase, we aim to deeply comprehend the code semantics present within each code commit. This is particularly crucial for discerning nuances related to security patches (details in~\Cref{sec:ft-cr}).

\subsection{Initial Retrieval via Hybrid Retriever}\label{sec:tfidf}
\begin{figure*}
    \centering
    \includegraphics[width=0.9\textwidth]{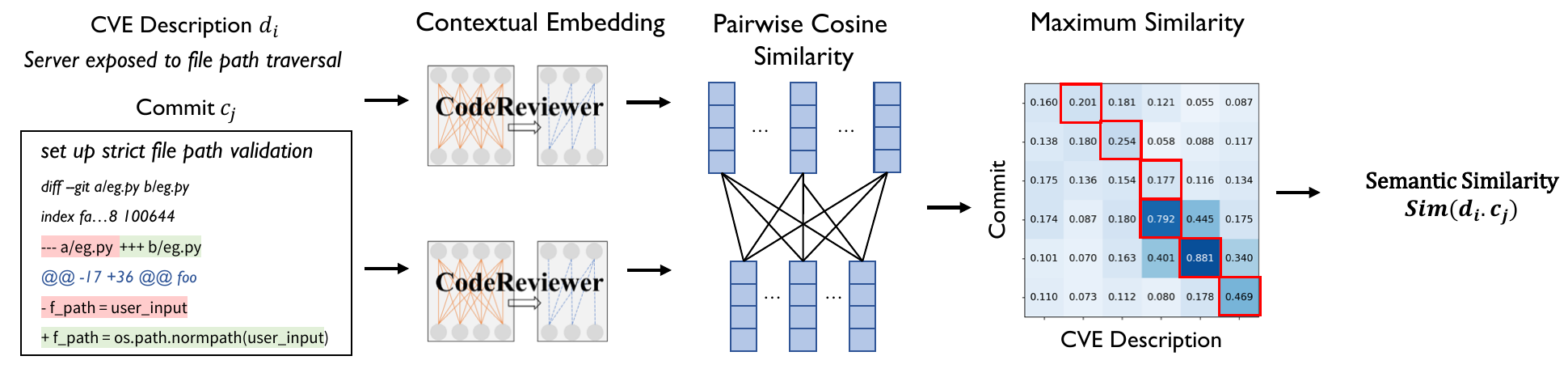}
    \caption{The workflow of our Semantic-based Retriever.}
    \label{fig:vector}
\end{figure*}

In the vast landscape of open-source repositories, developmental commits overwhelmingly outnumber security patches. To illustrate, the renowned Linux kernel project~\cite{linux-kernel} has amassed $1,215,313$ commits as of 15th September 2023. Yet, throughout its history, it has been associated with only $4,165$ CVEs~\cite{cve}. 
While existing ranking-based methods such as PatchScout \cite{Tan-ccs21} have made notable strides, they predominantly lean on handcrafted features to pinpoint security patches. Given the overwhelming number of commits, these methods might not fit well to consistently attain the desired precision.
To address this, we incorporate initial retrieval into the security patch tracing task.
Specifically, we utilize a hybrid approach to combine a lexical-based TF-IDF~\cite{chowdhury2010introduction} retriever and a semantic-based CodeReviewer (pretrained) retriever to take both lexical and semantic information into account. This is because, prior works~\cite{karpukhin-etal-2020-dense, zhang2020retrieval} show that sparse and dense retrievers can complement each other for more robust text retrieval. Due to the existence of large commits and the length constraint of CodeReviewer (maximum of 512 tokens), we preprocess diff files by extracting \textit{only the lines that involve code changes} and then limit the scope to the first 1,000 lines. The statistics show that it can get good coverage (98.6\%) of the patch samples on our dataset.

\subsubsection{Lexical-based Retriever.} 
TF-IDF~\cite{chowdhury2010introduction} stands out for its efficiency and its well-recognized capability to capture lexical similarities. At this stage, our objective is not to definitively locate the security patches but to considerably narrow down the pool of potential commits. By harnessing the capabilities of TF-IDF, we can effectively filter out commits less likely to be security patches, paving the way for a more in-depth analysis in the following stages of our approach. 

Formally, in our task, the term \( t \) represents individual words or tokens present in CVE descriptions or commits, which includes both commit messages and code diffs. Given a CVE description \(d_i \) \( \in \) \( \mathcal{D} \), both \(d_i \) and the corresponding commits \(c_j \in \mathcal{C}_i \) are treated as separate documents. The entire set of commits associated with a particular CVE, denoted as \( \mathcal{C}_i\), forms our corpus for \(d_i \).

The TF-IDF score for a term \( t \) in a document \( d \) (either a CVE description \( d_i \) or a commit \( c_j \)) within the corpus \( \mathcal{C}_i \) is given by:

\begin{equation}
\text{TF-IDF}(t, d, \mathcal{C}_i) = \text{TF}(t, d) \times \text{IDF}(t, \mathcal{C}_i)
\end{equation}
Here \( \text{TF}(t, d) \) is the term frequency of \( t \) in \( d \), calculated as the number of times \( t \) appears in \( d \) divided by the total number of terms in \( d \). \( \text{IDF}(t, \mathcal{C}_i) \) is the inverse document frequency of \( t \) in \( \mathcal{C}_i \), calculated as the logarithm of the total number of documents in \( \mathcal{C}_i \) divided by the number of documents containing \( t \).

To measure the similarity between the TF-IDF vectors of a given CVE description \( d_i \) and a code commit \( c^j \) ($c^j \in \mathcal{C}_i$), we employ cosine similarity between two vectors \( \vec{d_i} \) and \( \vec{c_j} \), which is defined as: 
\begin{equation}
    {cosine}(\vec{d_i}, \vec{c_j}) = \frac{\vec{d_i} \cdot \vec{c_j}}{\|\vec{d_i}\| \times \|\vec{c_j}\|}
\end{equation}
where \( \vec{d_i} \cdot \vec{c_j} \) is the dot product of the vectors, and \( \|\vec{d_i}\| \) and \( \|\vec{c_j}\| \) are the magnitudes of the vectors \( \vec{d_i} \) and \( \vec{c_j} \), respectively. 
The cosine similarity score ranges between 0 and 1, indicating the lexical similarity between the description and the commit. In this way, commits that are more similar to the given CVE description will have a higher cosine similarity score.

\subsubsection{Semantic-based Retriever.} 

Inspired by~\cite{zhang2019bertscore}, we adopt a pretrained CodeReviewer model to retrieve relevant patches by measuring their semantic similarity. Specifically, to encode the CVE description and commits, we use a CodeReviewer encoder to map each CVE description and commit pair $(d_i, c_j)$ (where $d_i \in \mathcal{D}$, $c_j \in \mathcal{C}_i$) to a fixed-size dense vector, leveraging its proficiency in analyzing code changes and understanding the semantics of defects such as vulnerabilities. 
Specifically, given a CVE description $d_i = <d_i^1, d_i^2, ..., d_i^{|d_i|}>$ and a candidate commit $c_j = <c_j^1, c_j^2, ..., c_j^{|c_j|}>$, we use contextual embeddings to represent the tokens and compute matching using cosine similarity (as shown in~\Cref{fig:vector}).

\noindent\textbf{Token Representation.} We use contextual embeddings to represent the tokens in the CVE description $d_i$ and candidate commit $c_j$, since its better semantic capturing when compared with word embeddings~\cite{zhang2019bertscore}. Contextual embeddings can generate different vector representations for the same word in different sentences depending on the surrounding words, which form the context of the target word. 
Specifically, We use a shared pretrained CodeReviewer (abbr. CRP) to separately encode the CVE description $d_i$ in $\mathcal{D}$ and each commits candidate $c_j$ in $\mathcal{C}_i$. We prepend a special token of \texttt{[CLS]} into its tokenized sequence and employ the final layer hidden state of the \texttt{[CLS]} token as the patch representation. We format each of the commits as $\{ [CLS], {diff}_j, [MSG], msg_j\}$.
Then the CVE description \(d_i\) and each commit \(c_j \in C_i\) are fed separately into the CRP encoder to obtain the sequences of token vectors, which can be formulated as:
$
S(d_i) = CRP_{\text{encoder}}(d_i)
$, and
$S(c_j) = CRP_{\text{encoder}}(msg_j; diff_j)$ respectively.

\noindent\textbf{Similarity Calculation.} The token representation facilitates a soft measure of similarity instead of
exact-string or heuristic matching in lexical-based methods. For each token vector in the CVE description, we denote them as $d_i^m\in S(d_i)$ and commit $d_j^n\in S(c_j)$, respectively. Then we calculate their cosine similarity to consider token relations between them.
To reduce the calculation cost to the inner product ${d_i^m}^T c_j^n$, we use pre-normalized vectors. While this measure considers tokens in isolation, the contextual embeddings contain information from the rest of the sentence. 

Based on this,  we calculate the complete score that matches each token in CVE description $d_i$ to a token in candidate commit $c_j$ to compute Recall and each token in $c_j$ to a token in $d_i$ to compute Precision. We use greedy matching to maximize the matching similarity score, where each token is matched to the most similar token in the other sentence. Finally, we combine precision and recall to compute an F1 measure. 
For a CVE description $d_i$ and its candidate commit $c_j$, the Recall (R), Precision (P), and F1 score (F1) are calculated as:
\begin{equation}
    R = \frac{1}{d_i} \sum_{d_i^m \in d_i}^{|d_i|} \max_{c_j^n \in c_j} {c_i^m}^T d_j^n
\end{equation}

\begin{equation}
    P = \frac{1}{c_j} \sum_{c_j^n \in c_j}^{|c_j|} \max {d_i^m}^T c_j^n
\end{equation}

\begin{equation}
   sim(d_i, c_j) = F1 = 2\frac{R \cdot P}{R+P}
\end{equation}

The F1 similarity score ranges between 0 and 1, indicating the semantic similarity $sim(d_i, c_j)$ between the given CVE description $d_i$ and the commit $c_j$. In this way, commits that are more semantically similar to the CVE description will have a higher F1 score.

\subsubsection{Hybrid Retriever.} 
To take both lexical and semantic information into account, we utilize a hybrid approach following~\cite{karpukhin-etal-2020-dense} to combine TF-IDF and CodeReviewer. 
The fusion of lexical and semantic similarities leverages their complementary analysis perspectives—lexical for word-based similarity and semantic for conceptual alignment (e.g., synonyms). 
Additionally, they share the same value space ([0,1]), facilitating straightforward additive fusion. The parameter $\lambda$ adjusts the emphasis on these features, allowing for a unified similarity metric. 
The similarity score is computed as $f_{\phi}(d_i, c_j)= sim(d_i, c_j) + \lambda \cdot cosine(\vec{d_i}, \vec{c_j})$, where $\lambda$ is a weight to balance the two retrievers. 
After conducting a parameter tuning process including a grid search over various values (from 0.1 to 10 with a step of 0.05), we found that $\lambda = 1$ in our experiment delivers optimal effectiveness among them. 
Nevertheless, we retain the parameter $\lambda$ to facilitate adaptation to different datasets.

Based on this combined similarity score, we rank the commits for a given CVE description. We retain the top-$k$ commits that have the highest similarity scores as candidates for the security patch. This yields a refined set of commits, which significantly narrows down the search space for locating the true patch. Indeed, a trade-off exists between efficiency and accuracy in this phrase. While there could be more accurate alternatives for retrieving these candidates such as BM25~\cite{bm25} and supervised dense retrieval approaches~\cite{colbert,karpukhin-etal-2020-dense}, the re-weighting and re-training process adds extra complexity. Fundamentally, we can further analyze the candidates and identify the patch through re-ranking. We provide the details of the re-ranking phrase in the next section.


\subsection{Re-ranking via Fine-tuning CodeReviewer}\label{sec:ft-cr}
As mentioned above, we have refined the list of commits to the top-$k$ most relevant candidates for each CVE. For this phase, we opt for CodeReviewer~\cite{codereview-fse22}, a state-of-the-art large language model pre-trained on the foundation of CodeT5~\cite{wang2021codet5}. There are two considerations for this choice.
\textbf{1) Encoder Specialization:} CodeReviewer's encoder is designed to deeply understand commit behaviors and issues, a feature not necessarily present or optimized in other models. This encoder specialization ensures that the model comprehends the intricate relationships between code changes and potential security implications, vital for matching commits to CVE descriptions. 
\textbf{2) Downstream Task Optimization:} Although our focus is not on generating code reviews, the fact that CodeReviewer's decoder is optimized for such downstream tasks indicates its ability to link code changes to descriptive text, a parallel to our objective of linking commits to CVE descriptions.
Given these advantages, we fine-tune CodeReviewer on the top-$k$ candidate commits, aiming to re-rank them based on their relevance to the respective CVE descriptions.


We only use the pre-trained encoder of CodeReviewer (abbr. CR) since our task can be basically viewed as a binary classification problem in the re-ranking phrase. Specifically, given a CVE description \(d_i\), and the top-\(k\) commits represented as \(C_k = \{(msg_j, diff_j)\}_{j=1}^{k}\), we format each commit as $\{ [CLS], diff_j, [MSG], msg_j\}$.
Then the CVE description \(d\) and each commit \(c_j \in C_k\) are encoded separately using the CR encoder to yield two sequences of vectors:
\begin{equation}
E(d) = CR_{\text{encoder}}(d)
\end{equation}
\begin{equation}
E(c_j) = CR_{\text{encoder}}(msg_j; diff_j)
\end{equation}
We obtain the vector representations of  \(d_i\) and \(c_j\) by extracting the hidden state in the last layer of the special token \texttt{[CLS]} at the beginning of $E(d_i)$ and $E(c_j)$ respectively.
The encoded vectors of the CVE description and the commit are concatenated:
\begin{equation}
V_j = [E(d_i); E(c_j)]
\end{equation}
We apply a linear classifier to the concatenated vector \(V_j\) for estimating the correlations:
\begin{equation}
y_j = \sigma(W \cdot V_j + b)
\end{equation}
where \(W\) is the weight matrix, \(b\) is the bias term, and \(\sigma\) denotes the sigmoid function ensuring the output lies between 0 and 1.

We utilize labeled data containing known CVE-commit pairs during the fine-tuning phase. The training goal is to minimize the binary cross-entropy loss:
\begin{equation}
\mathcal{L} = -\frac{1}{k} \sum_{j=1}^{k} [y_{\text{true},j} \log(y_j) + (1-y_{\text{true},j}) \log(1-y_j)]
\end{equation}
where \(y_{\text{true},j}\) is the ground truth label of the \(j^{th}\) sample, indicating whether commit \(c_j\) is related to the CVE description \(d_i\).

After fine-tuning, for any new CVE and set of commits, the model can compute the relevance scores. Commits can then be re-ranked based on the scores concerning the given CVE. 
This method ensures that the CR encoder understands both generic textual semantics and the specific indicators that tie a commit to the given CVE, making the approach specially tailored for our challenge.

\section{Evaluation}\label{sec:evaluation}

\subsection{Research Questions}

We aim to answer the following research questions:

\begin{itemize}[leftmargin=10pt]
    \item \textbf{RQ1: Effectiveness Analysis.} How effective is \ourtool when compared with existing baselines in tracing security patches?
    \item \textbf{RQ2: Ablation Analysis.} What impact does each component of \ourtool have on the overall performance?
    \item \textbf{RQ3: Distribution Analysis.} Does \ourtool exhibit notably high or low accuracy for certain vulnerability types or severity? 
    \item \textbf{RQ4: Practicality Analysis.} How effective is \ourtool in real-world applications, particularly when detecting security patches for CVEs without associated trace links?
\end{itemize}

\subsection{Dataset}

For the training and evaluation of our model, we compiled a dataset that encompasses both OSS vulnerabilities and their corresponding security patches. The dataset was assembled in two primary steps: 
\textbf{1)} \textbf{Initial Data Collection:} We began by collecting data from Wang et al.~\cite{wang2022vcmatch} and Tan et al.~\cite{Tan-ccs21}. We thereby obtained 1,669 unique CVEs \revised{from 10 OSS projects}.  
\textbf{2)} \textbf{Dataset Supplement:} To augment our dataset, we crawled vulnerabilities and their associated patches from multiple sources, including the official CVE, NVD, and Snyk~\cite{snyk} vulnerability databases. This initial collection yielded 3,585 unique CVEs \revised{from 532 OSS projects}. 
After eliminating duplicate entries, our finalized dataset comprises 4,789 unique CVEs and 4,870 distinct security patch commits, which involve \textbf{$532$} unique OSS projects, covering various programming languages including C/C++, Java, JavaScript, and PHP. This makes our dataset the most extensive collection of CVEs and security patches available to date.
Each dataset entry includes the CVE-ID along with its textual description and the corresponding security patch links. We also extracted related commits including commit messages and code diffs, primarily from GitHub~\cite{github} and GitLab~\cite{gitlab}. 

To create a robust training set, we followed the practices in prior works~\cite{Tan-ccs21, wang2022vcmatch} to sample 5,000 non-patch commits as negative samples for each CVE. However, in scenarios where a repository contained fewer than 5,000 commits in total, we included all available non-patch commits as negative samples. Finally, we got \textbf{21,781,044} commits in total. 
To the best of our knowledge, this is the largest dataset specific for the security patch tracing problem.

\subsection{Experiment Setup}\label{sec:setup}
We randomly split the 4,789 unique CVEs along with their corresponding commits in the proportion of $8: 1: 1$ to keep the same split settings as baselines~\cite{Tan-ccs21,wang2022vcmatch}. 
The maximum token length for CodeReviewer is set at 512, which represents its upper limit for processing capacity. Given the preprocessing of code diffs as detailed in Section~\ref{sec:approach}, this token length is actually sufficient for our purposes.
To preprocess data, we use the NLTK toolkit and the BPE tokenizer of CodeReviewer to tokenize CVE descriptions and commits. For the initial retrieval phase, we retrieve the top 100 commits from the initial 5,000 commits for each CVE. This threshold $k$ enables us to obtain a good balance between the coverage of true patches and noisy commits.
During fine-tuning, the batch size is set to $32$ and the maximum epoch is $20$. 
We adopt the widely-used Adam~\cite{kingma2017adam} as the optimizer with a learning rate of 5e-5 for training our model. All the above hyper-parameters are determined based on the validation set by selecting the best ones among some alternatives. 
All experiments ran on a server with $48$ CPU cores (Intel$^\circledR$ Xeon$^\circledR$ Silver 4214 CPU @ 2.20GHz), $252$ GB RAM, and 4 NVIDIA RTX 3090 GPUs ($24$ GB memory each).

\subsection{Baselines}
We benchmark our approach against state-of-the-art works in security patch tracing as presented in~\cite{Tan-ccs21,wang2022vcmatch}. \revised{Due to the absence of available replication packages of PatchScout~\cite{Tan-ccs21}, we implemented it independently by adhering to the default settings unless specified otherwise. }
Additionally, we consider two renowned baselines frequently used in the information retrieval domain for our evaluation: BM25~\cite{bm25}, a classic method for sparse retrieval~\cite{qiao2023optimizing}, and ColBERT~\cite{colbert}, known for dense  retrieval~\cite{karpukhin-etal-2020-dense}.

\subsection{Evaluation Metrics}
To ensure a fair comparison between \ourtool and the baselines, we utilize three metrics: Recall@K, Mean Reciprocal Rank (MRR)~\cite{gu2018deep}, and Manual Efforts@K. Recall@K and Manual Efforts@K have been previously employed in previous studies~\cite{wang2022vcmatch,Tan-ccs21}. We also incorporate MRR into our evaluation, given its established significance in ranking systems. 

\subsubsection{Recall@K} 
Recall@$K$ refers to the ratio of the number of security patches traced in the top-K results to the number of all security patches for a given $K$. Hence, a higher Recall@K score means better performance.

\subsubsection{Mean Reciprocal Rank (MRR)} MRR is a widely used evaluation metric for ranking systems, particularly in the domain of information retrieval. It emphasizes the importance of the position of the first relevant result in a list of retrieved documents, making it especially relevant for security patch tracing where we typically seek a single commit. MRR is defined mathematically as follows: 
\begin{equation}
MRR = \frac{1}{|{D}|} \sum_{i=1}^{|{D}|} \frac{1}{\text{rank}_i}
\end{equation}

In this equation, $|{D}|$ represents the total number of CVEs, and $\text{rank}_i$ denotes the position of the first security patch for the $i$-th CVE. The MRR values range between 0 and 1, with higher values indicating better retrieval performance. By considering the inverse of the rank of the first relevant result, MRR encourages systems to prioritize the most relevant information at the top of the list, thus improving user satisfaction and system efficiency. 
The higher the MRR value, the better the security patch tracing performance.

\subsubsection{Manual Efforts@K}
In the pursuit of tracing security patches within OSS, the Manual Efforts@K metric emerges as a classic metric. It represents the manual inspection effort required to locate the correct patch within the top-K results. If the desired security patch is found within these results, the effort corresponds to its rank. However, if the patch is not within the top-K, the effort is $K$, indicating a comprehensive search without success. 
Drawing from related work~\cite{Tan-ccs21}, the metric is mathematically expressed as:
\begin{equation}
    \text{Manual Efforts@K} = \frac{\sum_{i=1}^{n} \min(\text{rank}_i, K)}{n}
\end{equation}

A lower Manual Effort@K score is indicative of a more efficient and effective method for tracing security patch commits. This aids NVD security experts in mitigating the extensive manual work associated with tracing security patches, reducing inspection time, and enhancing patch detection accuracy. 

\section{Results and Discussion}\label{sec:results}
We investigate the following research questions to provide a thorough analysis of the experimental results.

\subsection{RQ1: Effectiveness Analysis} 

\begin{table}[]
\caption{The effectiveness of \ourtool and baselines to trace patch commits.}
\label{tab:effectivess_comparison}
\resizebox{0.99\columnwidth}{!}{
\begin{tabular}{cccccc}
\hline
\textbf{Recall@K} & \textbf{PatchScout} & \textbf{VCMatch} & \textbf{BM25} & \textbf{ColBERT} & \textbf{\ourtool} \\ \hline
\rowcolor[HTML]{EFEFEF} 
\cellcolor[HTML]{EFEFEF}\textbf{K=1} & \cellcolor[HTML]{EFEFEF}46.25\% & 55.93\% & 11.88\% & 26.29\% & \textbf{79.23\%} \\
\textbf{K=2}                         & 48.51\%                         & 57.72\% & 16.88\% & 31.49\% & \textbf{79.30\%} \\
\rowcolor[HTML]{EFEFEF} 
\cellcolor[HTML]{EFEFEF}\textbf{K=3} & \cellcolor[HTML]{EFEFEF}48.72\% & 58.07\% & 19.58\% & 34.41\% & \textbf{79.57\%} \\
\textbf{K=4}                         & 48.72\%                         & 58.42\% & 20.83\% & 37.23\% & \textbf{79.64\%} \\
\rowcolor[HTML]{EFEFEF} 
\cellcolor[HTML]{EFEFEF}\textbf{K=5} & \cellcolor[HTML]{EFEFEF}48.92\% & 59.58\% & 22.08\% & 38.38\% & \textbf{79.91\%} \\
\textbf{K=6}                         & 48.92\%                         & 61.88\% & 22.50\% & 40.93\% & \textbf{79.97\%} \\
\rowcolor[HTML]{EFEFEF} 
\cellcolor[HTML]{EFEFEF}\textbf{K=7} & \cellcolor[HTML]{EFEFEF}48.92\% & 63.42\% & 23.54\% & 41.96\% & \textbf{80.04\%} \\
\textbf{K=8}                         & 48.92\%                         & 63.42\% & 25.00\% & 43.22\% & \textbf{80.31\%} \\
\rowcolor[HTML]{EFEFEF} 
\cellcolor[HTML]{EFEFEF}\textbf{K=9} & \cellcolor[HTML]{EFEFEF}48.92\% & 63.42\% & 25.83\% & 44.45\% & \textbf{80.35\%} \\
\textbf{K=10}                        & 48.92\%                         & 63.42\% & 26.04\% & 44.95\% & \textbf{80.63\%} \\
\hline
\hline
\textbf{MRR}                         & 0.3824                          & 0.6195  & 0.1736  & 0.3240  & \textbf{0.7951}  \\ \hline
\end{tabular}
}
\end{table}

\begin{table*}[]
\caption{Manual Efforts@K (ME@K) of \ourtool and baselines to trace security patch commits.}
\label{tab:manualefforts_comparison}
\resizebox{0.9\textwidth}{!}{
\begin{tabular}{cccccc|cccccc}
\hline
\textbf{ME@K} & \textbf{PatchScout} & \textbf{VCMatch} & \textbf{BM25} & \multicolumn{1}{l}{\textbf{ColBERT}} & \textbf{\ourtool} & \textbf{ME@K} & \textbf{PatchScout} & \textbf{VCMatch} & \textbf{BM25} & \multicolumn{1}{l}{\textbf{ColBERT}} & \textbf{\ourtool} \\ \hline
\rowcolor[HTML]{EFEFEF} 
\textbf{K=1} & 1.00 & 1.00 & 1.00 & 1.00 & \textbf{1.00} & \textbf{K=8} & 4.61 & 4.47 & 6.29 & 5.38 & \textbf{2.38} \\
\textbf{K=2} & 1.54 & 1.51 & 1.86 & 1.71 & \textbf{1.20} & \textbf{K=9} & 5.12 & 4.51 & 6.95 & 5.94 & \textbf{2.58} \\
\rowcolor[HTML]{EFEFEF} 
\textbf{K=3} & 2.05 & 1.54 & 2.68 & 2.38 & \textbf{1.40} & \textbf{K=10} & 5.63 & 5.38 & 7.64 & 6.49 & \textbf{2.77} \\
\textbf{K=4} & 2.57 & 2.28 & 3.46 & 3.01 & \textbf{1.60} & \textbf{K=20} & 10.59 & 10.14 & 13.67 & 11.70 & \textbf{4.71} \\
\rowcolor[HTML]{EFEFEF} 
\textbf{K=5} & 3.08 & 2.35 & 4.23 & 3.63 & \textbf{1.80} & \textbf{K=30} & 14.75 & 13.91 & 19.43 & 16.52 & \textbf{6.65} \\
\textbf{K=6} & 3.59 & 3.43 & 4.94 & 4.23 & \textbf{1.99} & \textbf{K=50} & 29.46 & 26.01 & 29.69 & 25.26 & \textbf{10.52} \\
\rowcolor[HTML]{EFEFEF} 
\textbf{K=7} & 4.10 & 3.82 & 5.60 & 4.81 & \textbf{2.19} & \textbf{K=100} & 41.86 & 34.47 & 50.91 & 44.92 & \textbf{20.21} \\ \hline
\end{tabular}
}
\end{table*}

Table~\ref{tab:effectivess_comparison} and Table~\ref{tab:manualefforts_comparison} show the effectiveness of different approaches in terms of Recall@K, MRR, and Manual Efforts@K, with the best one of each metric marked in bold. 
Table~\ref{tab:effectivess_comparison} reveals that \ourtool significantly outperforms all the SOTA approaches across different values of \(K\), for the Recall@K metric. The superiority of \ourtool is most prominent at \(K=1\) where it achieves a Recall of 79.23\%, markedly higher than PatchScout's 46.25\%, VCMatch's 55.93\%, BM25 's 11.88\%, and ColBERT's 26.29\%. This trend continues as \(K\) increases, showcasing the consistent effectiveness of \ourtool in locating security patch commits within the top-$K$ results. 
Notably, the Recall@K for \ourtool remains above 79\% for all values of \(K\), highlighting its robustness in tracing relevant security patches. 
The MRR further confirms the effectiveness of \ourtool with a score of \mrr, significantly outpacing the $0.6195$ and $0.3824$ attained by VCMatch and PatchScout, respectively. 
This superior performance can be attributed to our two-phase approach that combines the lexical-level understanding from TF-IDF with the semantic understanding from the fine-tuned CR model. 
As shown in Table~\ref{tab:manualefforts_comparison}, \ourtool requires much less manual effort compared to other methods. 
At \(K=1\), all methods tie with a score of 1.00, indicating minimal manual effort required regardless of the existence of true patches. However, as \(K\) increases, \ourtool consistently requires less manual effort compared to the other baselines. For instance, at \(K=10\), \ourtool registers a score of $2.77$, which is considerably less than the manual effort demanded by the best baseline. This trend underscores the efficiency of \ourtool, particularly as the value of \(K\) rises, demonstrating a lower manual effort requirement for practitioners aiming to trace security patches. 
\revised{Meanwhile, the observed lower effectiveness of PatchScout and VCMatch on our dataset compared to their published results~\cite{Tan-ccs21, wang2022vcmatch} can be attributed to two factors: 
\ding{172} \textit{Data diversity:} As discussed before, PatchScout and VCMatch rely on handcrafted word-based similarities derived from their original dataset, which diminishes with increased data diversity. Such features struggle to capture relevant patch characteristics in a varied dataset. Our expanded dataset, featuring 4,789 CVEs from 532 OSS projects, presents an apparent contrast to their original dataset's 658 (1,669 for VCMatch) CVEs from only 5 (10 for VCMatch) OSS projects. This significant increase in both the number of CVEs and the diversity of originating projects introduces greater complexity, challenging their ability to accurately identify patches. 
\ding{173} \textit{Language Specificity:} Their design primarily focuses on C/C++ OSS projects, which may not generalize well to other languages like Java and PHP, etc. 
These factors collectively lead to the variance in effectiveness between the original and our new datasets when applying PatchScout and VCMatch. Moreover, this comparison underscores an additional advantage of \ourtool: unlike PatchScout and VCMatch, \ourtool does not rely on predefined, language-specific features. This design enhances \ourtool's scalability and applicability across various programming languages. 
}

The observed results underscore the efficacy of \ourtool in tracing security patch commits compared to existing baseline approaches. 
While previous ranking-based methods including PatchScout and VCMatch have made significant contributions when compared with match-based approaches~\cite{Tan-ccs21,wang2022vcmatch}, their reliance on \textit{handcrafted features} (predominantly \textit{lexical only}). However, in practice, CVE descriptions and commits often use different terminologies to describe the same vulnerability including synonyms or varying phrasings. This discrepancy becomes particularly challenging when CVE descriptions or commits do not adhere to high-quality documentation standards, a situation frequently encountered with CVEs lacking associated patches. 
Consequently, their performance proves to be inadequate on the large and diverse projects that involve arbitrary descriptions and commits.

In contrast, \ourtool leverages TF-IDF and CR to understand both the lexical and semantic aspects of the CVE description and commits, thereby achieving higher accuracy. Moreover, the fine-tuning process of CR allows our model to adapt to the specific semantics and patterns commonly found in security patches. This adaptability is a marked edge over methods such as PatchScout, which cannot adjust to the data on which they are deployed.

\begin{table}[t]
\centering
\caption{\revised{Efficiency of PatchScout, VCMatch, and \ourtool.}}
\label{tab:efficiency}
\resizebox{0.9\columnwidth}{!}{
\begin{tabular}{BBBB}
\toprule
\textbf{Tool} & \textbf{PatchScout} & \textbf{VCMatch} & \textbf{PatchFinder} \\ \midrule
Time Cost per CVE (s) & 41.57 & 43.35 & 46.83 \\ \bottomrule
\end{tabular}
}
\end{table}

\noindent\revised{\textbf{Efficiency Analysis.}}
\revised{
While \ourtool employs a two-phase approach and involves fine-tuning CR, it maintains computational efficiency. 
To demonstrate this, we compared \ourtool with PatchScout~\cite{Tan-ccs21} and VCMatch~\cite{wang2022vcmatch} across our test set of 480 CVEs, conducting three trials to ensure accuracy. 
The results displayed in~\Cref{tab:efficiency} reflect that \ourtool's efficiency is closely competitive with the baselines. Despite a slightly higher overall time cost (3.48-5.26s), \ourtool's performance is competitive, especially when considering its enhanced accuracy and scalability. 
Notably, Phase-2 required only 1.1 seconds on average per CVE. 
The initial retrieval phase consumes most of the time due to the extraction of lexical and semantic features. Specifically, the time cost for Phase-1 is 45.75 seconds per CVE. 
To speed it up, \ourtool's efficiency can be easily enhanced by applying a vector database for retrieval, such as Faiss~\cite{faiss} or Redis~\cite{Redis} in this phase. }



\smallskip
\noindent\fbox{
	\parbox{0.95\linewidth}{
		\textbf{Answer to RQ1:} \textit{Compared with the four baselines, \ourtool demonstrated superior effectiveness, achieving improvements of  $17.42\%$-$54.59\%$ in Recall@10, $0.1756$-$0.6215$ in MRR, and effectively reducing Manual Efforts@100 by $14.26$-$30.7$. These results underscore the effectiveness of \ourtool in patch tracing.}}
}

\subsection{RQ2: Ablation Analysis} 

\revised{In our proposed approach, we integrate two primary components: an initial retrieval using a hybrid retriever consisting of TF-IDF and pre-trained CodeReviewer (Phase-1), followed by a subsequent re-ranking using a fine-tuned CodeReviewer model (Phase-2). The input to our system encompasses the CVE description, commit messages, and code diffs. 
To dissect their effectiveness, we conducted a detailed ablation study by examining the performance of \ourtool under various configurations: 
\ding{172} Using only the lexical-based retriever, i.e., our TF-IDF retriever for tracing security patches (termed \textbf{``TF-IDF''}), 
\ding{173} Using only the semantic-based retriever, i.e., pre-trained CodeReviewer model for tracing security patches (termed \textbf{``CR$_{pretrain}$''}), 
\ding{174} Utilizing only the CVE description and \textit{commit messages} (termed \textbf{``Msg''}) for security patch tracing, and \ding{175} Utilizing only CVE description and \textit{code diffs} (termed \textbf{``Diff''}) for tracing, 
\ding{176} Employing the initial retrieval phase only (termed \textbf{``Phase-1''}), and 
\ding{177} Directly fine-tuning CodeReviewer without the first phase (termed \textbf{``Phase-2''}). 
}
\begin{table}[t]
\centering
\caption{Contribution of individual components in \ourtool in terms of Recall@K.}
\label{tab:ablation_recall}
\resizebox{1\columnwidth}{!}{
\begin{tabular}{cccccBBc}
\hline
\textbf{Recall@K} & \textbf{TD-IDF} & \textbf{CR$_{pretrain}$} & \textbf{Diff} & \textbf{Msg} & \textbf{Phase-1} & \textbf{Phase-2} & \textbf{\ourtool} \\ \hline
\rowcolor[HTML]{EFEFEF} 
\textbf{K=1} & 35.21\% & 28.75\% & \cellcolor[HTML]{EFEFEF}35.26\% & \cellcolor[HTML]{EFEFEF}63.85\% & 41.04\% & 0.21\% & \cellcolor[HTML]{EFEFEF}\textbf{79.23\%} \\
\textbf{K=2} & 40.21\% & 32.71\% & 42.60\% & 63.96\% & 47.71\% & 0.42\% & \textbf{79.30\%} \\
\rowcolor[HTML]{EFEFEF} 
\textbf{K=3} & 45.21\% & 34.58\% & \cellcolor[HTML]{EFEFEF}45.36\% & \cellcolor[HTML]{EFEFEF}64.06\% & 51.88\% & 0.42\% & \cellcolor[HTML]{EFEFEF}\textbf{79.57\%} \\
\textbf{K=4} & 47.71\% & 36.67\% & 46.67\% & 64.17\% & 54.38\% & 1.25\% & \textbf{79.64\%} \\
\rowcolor[HTML]{EFEFEF} 
\textbf{K=5} & 51.04\% & 38.33\% & \cellcolor[HTML]{EFEFEF}48.80\% & \cellcolor[HTML]{EFEFEF}64.27\% & 56.04\% & 1.46\% & \cellcolor[HTML]{EFEFEF}\textbf{79.91\%} \\
\textbf{K=6} & 52.71\% & 39.79\% & 49.27\% & 64.38\% & 57.50\% & 1.46\% & \textbf{79.97\%} \\
\rowcolor[HTML]{EFEFEF} 
\textbf{K=7} & 53.75\% & 40.21\% & \cellcolor[HTML]{EFEFEF}50.57\% & \cellcolor[HTML]{EFEFEF}64.48\% & 58.54\% & 1.46\% & \cellcolor[HTML]{EFEFEF}\textbf{80.04\%} \\
\textbf{K=8} & 54.79\% & 41.88\% & 52.08\% & 64.58\% & 59.17\% & 1.46\% & \textbf{80.31\%} \\
\rowcolor[HTML]{EFEFEF} 
\textbf{K=9} & 56.25\% & 43.13\% & \cellcolor[HTML]{EFEFEF}52.71\% & \cellcolor[HTML]{EFEFEF}64.58\% & 60.83\% & 1.46\% & \cellcolor[HTML]{EFEFEF}\textbf{80.35\%} \\
\textbf{K=10} & 57.29\% & 43.96\% & 55.70\% & 64.58\% & 61.46\% & 1.88\% & \textbf{80.63\%} \\ \hline
\hline
\textbf{MRR} & 0.4243 & 0.3394 & 0.4146 & 0.6403 & 0.4827 & -5.87E-06 & \textbf{0.7951} \\ \hline
\end{tabular}
}
\end{table}

As shown in Table~\ref{tab:ablation_recall} and Table~\ref{tab:ablation_me}, the substantial improvement in terms of all metrics demonstrates the superiority of our two-phase approach over either a lexical-based or semantic-based retriever. \ourtool attains a Recall@1 of 79.23\%, which is more than double the performance when solely relying on TF-IDF or pretrained CodeReviewer. Notably, CR$_{pretrain}$, which has not undergone fine-tuning, still manages an acceptable score, especially when compared with ColBERT as shown in~\Cref{tab:effectivess_comparison}. This underscores the importance of domain-specific LLMs (CodeReviewer in \ourtool) in understanding the underlying semantics in this task. 

Similarly, both commit messages and code diffs play pivotal roles in capturing the nuanced semantics of commits, enabling accurate security patch tracing. 
\revised{
The comparison suggests that commit messages have a good positive impact. 
This is likely because both CVE descriptions and commit messages are written in natural language, whereas code diffs are in various programming languages. This makes their lexical structures much less aligned, and thus relying solely on diffs in the initial phase can result in decreased recall. 
Nonetheless, commit messages and code diffs serve complementary roles in patch tracing within \ourtool, rather than being mutually exclusive. Table~\ref{tab:ablation_recall} and~\ref{tab:ablation_me} indicate that incorporating code diffs alongside commit messages significantly boosts the \ourtool's effectiveness, demonstrating a notable improvement in Recall@1 from 63.85\% to over 79.23\% at Recall@1, an enhancement of 15.38\%. 
It contributes to tracing an additional $16.05\%$ of security patches that are untraceable by commit messages alone, and enhances the MRR by $0.1548$, thereby reducing the manual effort by $5.95$ commits for $K=100$. 
This underlines the importance of code diffs in enriching our semantic analysis for more accurate patch identification. }

Interestingly, based on our results in Table~\ref{tab:effectivess_comparison} and ~\ref{tab:manualefforts_comparison}, TF-IDF tends to outperform BM25 for retrieving the commits that cover true patches.
As explored in~\cite{kadhim2019term}, BM25 employs a term saturation model and document length normalization for long texts. In our context, these aspects might inadvertently prioritize exact matches over partial ones, making it less effective for the concise nature of CVE descriptions and commits. Hence, weighing both effectiveness and efficiency, TF-IDF emerges as our preferred choice for this task. 

\revised{
Moreover, the quantitative results presented in Tables~\ref{tab:ablation_recall} and~\ref{tab:ablation_me} clarify our motivation for adopting a two-phase design and demonstrate its essential role in addressing the challenges previously mentioned. Specifically, the \textbf{Phase-1} configuration effectively narrows down the dataset, ensuring the LLM in the re-ranking phase to focus its analysis on a more refined set of candidates. 
This is evident from its Recall@K, which reaches up to 61.46\% at K=10, and an MRR of 0.4827, underscoring its critical contribution to ensuring broad coverage. 
Conversely, when the analysis is conducted with only \textbf{Phase-2}, its performance significantly drops with Recall@K peaking at merely 1.88\% at K=10. This obvious under-performance highlights the LLM's challenges in dealing with vast, imbalanced datasets (with a patch to non-patch ratio of 1:5000), thereby illustrating the necessity of the initial retrieval phase in refining the dataset for subsequent LLM re-ranking. 
This refinement shifts the ratio from 1:5000 to a more manageable 1:100, ensuring focused and effective analysis.
}

\smallskip
\noindent\fbox{
	\parbox{0.95\linewidth}{
		\textbf{Answer to RQ2:} \textit{\revised{Both main components and input of \ourtool effectively contribute to the overall performance. 
        Commit messages are important, while code diffs further complement and enhance semantic understanding. 
        Meanwhile, the two-phase design of PatchFinder proves to be crucial for the overall effectiveness in locating patches, where only using one of the two phases can result in a significant decline in performance.}
  } 
	}
}

\begin{table}[t]
\centering
\caption{Contribution of individual components in \ourtool in terms of Manual Efforts@K (ME@K).}
\label{tab:ablation_me}
\resizebox{0.99\columnwidth}{!}{
\begin{tabular}{cccccBBc}
\hline
{\textbf{ME@K}} & {\textbf{TF-IDF}} & {\textbf{CR$_{pretrain}$}} & {\textbf{Diff}} & {\textbf{Msg}} & {\textbf{Phase-1}} & {\textbf{Phase-2}} & {\textbf{\ourtool}} \\ \hline
\rowcolor[HTML]{EFEFEF} 
\textbf{K=1} & 1.00 & 1.00 & 1.00 & \cellcolor[HTML]{EFEFEF}1.00 & 1.00 & 1.00 & \textbf{1.00} \\
\textbf{K=2} & 1.54 & 1.71 & 1.51 & 1.25 & 1.59 & 2.00 & \textbf{1.20} \\
\rowcolor[HTML]{EFEFEF} 
\textbf{K=3} & 2.05 & 2.38 & 1.54 & \cellcolor[HTML]{EFEFEF}1.51 & 2.11 & 2.99 & \textbf{1.40} \\
\textbf{K=4} & 2.57 & 3.04 & 2.28 & 1.76 & 2.59 & 3.99 & \textbf{1.60} \\
\rowcolor[HTML]{EFEFEF} 
\textbf{K=5} & 3.08 & 3.67 & 2.35 & \cellcolor[HTML]{EFEFEF}2.02 & 3.05 & 4.98 & \textbf{1.80} \\
\textbf{K=6} & 3.59 & 4.29 & 3.43 & 2.27 & 3.49 & 5.96 & \textbf{1.99} \\
\rowcolor[HTML]{EFEFEF} 
\textbf{K=7} & 4.10 & 4.89 & 3.82 & \cellcolor[HTML]{EFEFEF}2.53 & 3.91 & 6.95 & \textbf{2.19} \\
\textbf{K=8} & 4.74 & 5.49 & 4.61 & 2.78 & 4.33 & 7.93 & \textbf{2.38} \\
\rowcolor[HTML]{EFEFEF} 
\textbf{K=9} & 5.19 & 6.07 & 5.12 & 3.03 & 4.73 & 8.92 & \textbf{2.58} \\
\textbf{K=10} & 5.63 & 6.64 & 5.63 & 3.29 & 5.13 & 9.90 & \textbf{2.77} \\
\rowcolor[HTML]{EFEFEF} 
\textbf{K=20} & 9.52 & 11.94 & 10.59 & 5.83 & 8.63 & 19.67 & \textbf{4.71} \\
\textbf{K=30} & 12.76 & 16.57 & 14.75 & 8.37 & 11.66 & 29.20 & \textbf{6.65} \\
\rowcolor[HTML]{EFEFEF} 
\textbf{K=50} & 18.58 & 24.73 & 29.46 & 13.45 & 16.96 & 47.81 & \textbf{10.52} \\
\textbf{K=100} & 31.19 & 43.02 & 41.86 & 26.16 & 27.90 & 93.07 & \textbf{20.21} \\ \hline
\end{tabular}
}
\end{table}

\subsection{RQ3: Distribution Analysis}
This RQ focuses on a detailed investigation of the outcomes from \ourtool. The primary aim is to explore if there exists a correlation between the types of vulnerabilities and the true patches identified by \ourtool. 
Specifically, we are interested in examining whether \ourtool's retrieval accuracy varies across different vulnerability categories and their respective severity levels.

To conduct this analysis, we categorized the vulnerabilities in our test dataset based on their Common Weakness Enumeration (CWE) identifiers~\cite{cwe} and Common Vulnerability Scoring System (CVSS) V2 scores~\cite{cvss2}. 
Our analysis reveals that \ourtool is particularly effective in tracing security patches for specific types of vulnerabilities. It shows exceptional performance for vulnerabilities categorized under \textit{CWE-125: Out-of-bounds Read} and \textit{CWE-119: Improper Restriction of Operations within the Bounds of a Memory Buffer}, achieving a tracing success rate of 100\% (46/46) and 78.85\% (41/52), respectively. While \ourtool generally performs well across various CWE types, it does exhibit lower effectiveness for \textit{CWE-122: Heap-based Buffer Overflow} and \textit{CWE-834: Excessive Iteration}, with tracing ratios of 20\% (1/5) and 33.3\% (1/3), respectively. This indicates that the vulnerability distribution does have a significant impact on the effectiveness. 
The lower effectiveness observed for certain CWE types, such as CWE-122 and CWE-834, might be attributed to the inherent complexity of these vulnerabilities. For instance, heap-based buffer overflows (CWE-122) can manifest in various ways in the code, making them harder to trace even though using CR. Addressing this challenge might require more specialized features tailored to specific vulnerability types.
Additionally, gathering more training data related to these challenging CWEs could enhance the model's understanding and improve performance. 

We further investigated the relationship between the severity of vulnerabilities, as indicated by their CVSS V2 scores, and the level of difficulty in tracing their corresponding security patches. Our analysis reveals a notable correlation: vulnerabilities with higher CVSS scores are generally easier to trace. Specifically, \ourtool successfully traced 84.52\% (71/84) of high-severity CVEs. Surprisingly, the tracing success rate for medium-severity CVEs was slightly higher, at 86.09\% (192/223). However, the tracing success rate for low-severity CVEs was the lowest, at 78.03\% (135/173). One plausible explanation for this trend could be that high and medium-severity vulnerabilities often come with detailed descriptions, immediate developer attention, and increased community scrutiny, all of which facilitate more accurate tracing. In contrast, low-severity CVEs often receive less detailed documentation and lower levels of developer and community focus. Besides, the patches written by security analysts can differ according to the severity of vulnerabilities, which need dedicated strategies to capture their semantics. These findings highlight that the vulnerability severity involving the quality and characteristics of CVE artifacts is a non-negligible factor to consider for further refinements.


\smallskip
\noindent\fbox{
\parbox{0.95\linewidth}{
\textbf{Answer to RQ3:} \textit{The vulnerability distribution does have a significant impact on the effectiveness. The performance is also tied to the severity of vulnerabilities, showing better outcomes for high and medium-severity vulnerabilities.  Implementing specialized strategies and utilizing data augmentation could improve outcomes for more difficult types and vulnerabilities of low severity.}
}
}


\subsection{RQ4: Practicality Analysis} 
In evaluating the practical utility of \ourtool, we initially curated a dataset of $212,074$ CVE entries from NVD as of April 2023. However, we encountered a significant challenge in accurately identifying CVEs without patches since the ``patch'' tags in the NVD are often imprecise~\cite{Tan-ccs21}: some entries with patches lack the corresponding tag, while others may be inaccurately tagged as having a patch but in fact, they do not.\footnote{There are 58.28\% CVEs ($123,587$/$212,074$) missing ``patch'' tags in our initial dataset.} Hence, we opted to focus on a more reliable subset: CVEs affecting OSS and known to lack patches. 
To this end, we leveraged the OSS project list maintained by OSS-Fuzz~\cite{ossfuzzp90:online}, resulting in 1,199 OSS projects. 
Further refining our selection, we excluded CVEs with any commit links on NVD, indicating the potential presence of a patch. This meticulous process yielded a targeted set of $473$ CVEs, belonging to 268 OSS projects. 

Upon deploying \ourtool on this curated set, we derived the top-10 ranked outputs for each CVE. 
From this pool, we initially manually reviewed and traced \numpatch patches. 
\revised{These patches achieved a ranking of 1.65 in \ourtool’s output on average. The entire review process was efficiently conducted, taking a total of $13.31$ man-hours. We then submitted these patches to CNAs for review.
Notably, \numconfirm of these were subsequently confirmed by CNAs so far~\cite{website}. }
This achievement underscores the tangible benefits of \ourtool and its potential to uncover and address omissions in current vulnerability databases. 

\smallskip
\noindent\fbox{
	\parbox{0.95\linewidth}{
		\textbf{Answer to RQ4:} \textit{\ourtool effectively traces missing security patches for a significant set of CVEs in the NVD. From a curated set of 473 CVEs, \ourtool's top-10 ranked outputs led to the identification of numerous plausible patch commits. Of the \numpatch manually reviewed patches (averaging a rank of 1.65), \numconfirm were confirmed by CNAs, underscoring \ourtool's practicality and its ability to address gaps in current vulnerability databases.
	    }
    }
}


\subsection{Case Study}
Among the \numpatch patches we traced in RQ4, a particularly illustrative instance is CVE-2022-31814 from the ``pfSense-pfBlockerNG'' project. As shown in~\Cref{lst:case_study}, this vulnerability allows remote attackers to execute arbitrary OS commands via specific manipulations. The associated patch commit, with its seemingly innocuous commit message ``Update index.php'' does not directly indicate its relevance to the CVE. This ambiguity poses challenges for tools like PatchScout and VCMatch who failed to recognize it. Their reliance on direct textual correlations (such as ``\# shared file names'', ``\# shared function names'', and ``\# shared words'', etc.) between CVE descriptions and commits can be limited, especially when descriptions and commits employ synonyms or different phrasings.

In contrast, \ourtool outperforms in such situations since harnessing the strengths of TF-IDF and CodeReviewer. Unlike other tools, it effectively deals with commits having limited information. Specifically, the patch commit in~\Cref{lst:case_study} initially ranked 47th in our lexical-based retriever due to its brief message. However, the semantic-based retriever recognized its relevance, where the addition of \texttt{escapeshellarg} at Lines 23-24 crucially sanitizes shell metacharacters in the HTTP Host header, directly addressing the vulnerability. Such intricate changes, often missed by other tools, are accurately identified by \ourtool due to its outstanding semantic analysis of code and text. As a result, our hybrid retriever improved its rank to 23rd. In the subsequent phase, the top-100 results, including this commit, were analyzed further. Here, the fine-tuned CR, adept at understanding code semantics, elevated its rank to 7th, placing it within the top-10 results. 

\subsection{\revised{Discussion on False Negatives}}
\revised{
We further analyzed the missed patch commits of 95 CVEs and summarized them into three main causes. 
\begin{itemize}[leftmargin=12pt]
    \item \textbf{Low-Quality CVE Descriptions (28/95):} Some CVE descriptions lack sufficient detail for effective patch tracing. Notably, CVE-2022-0080~\cite{CVE-2022-0080} only mentions ``mruby is vulnerable to Heap-based Buffer Overflow'', missing any vulnerability information except for vulnerability type. 
    \item \textbf{Giant Commits (41/95):}
            \ding{172} \textit{Irrelevant File Changes:} Commits such as~\cite{patch-CVE-2021-37686} (a patch commit for CVE-2021-37686) often include changes unrelated to the patch, such as refactoring, which may hinder the patch's intent by introducing the noise. 
            \ding{173} \textit{Token Limit Exceedance:} Some patch commits (e.g., \cite{patch-CVE-2017-18922} for CVE-2017-18922) exceed CodeReviewer's token limit, affecting \ourtool's ability to analyze them fully, even though we have pruned the code diffs before feeding them into CodeReviewer (as discussed in Section~\ref{sec:tfidf}). 
    \item \textbf{Confusing Commits (26/95):} Certain commits such as~\cite{CVE-2017-13146-confusing} deliberately obscure their patching role (fixing CVE-2017-13146 in this case). Still, non-patch commits within the same repository conversely claim it ``fix'' something, making it challenging for our hybrid retriever to identify candidate patches accurately. 
\end{itemize}
}

\begin{lstlisting}[language=diff,caption=The patch commit for CVE-2022-31814.,label={lst:case_study}, float={t}]
CVE Description:
pfSense pfBlockerNG through 2.1.4_26 allows remote attackers to execute arbitrary OS commands as root via shell metacharacters in the HTTP Host header. NOTE: 3.x is unaffected.
****************************************************************
commit 071bdcf2d918c3e51cde11cf81fbd9b6f0379d7e
Author: BBcan177 <bbcan177@gmail.com>
Date:   Sun Jun 5 13:25:24 2022 -0400

    Update index.php

diff --git a/net/pfSense-pkg-pfBlockerNG/files/usr/local/www/pfblockerng/www/index.php 
           b/net/pfSense-pkg-pfBlockerNG/files/usr/local/www/pfblockerng/www/index.php
index 8b8af0fab6b8..63f898b89246 100644
--- a/net/pfSense-pkg-pfBlockerNG/files/usr/local/www/pfblockerng
---/www/index.php
+++ b/net/pfSense-pkg-pfBlockerNG/files/usr/local/www/pfblockerng
+++/www/index.php
@@ -48,7 +48,7 @@ if (!empty($log)) {

 // Query DNSBL Alias for Domain List.
 $query = str_replace('.', '\.', htmlspecialchars($_SERVER['HTTP_HOST']));
-exec("/usr/bin/grep -l ' \"{$query} 60 IN A' /var/db/pfblockerng
-/dnsblalias/*", $match);
+exec("/usr/bin/grep -l " . escapeshellarg("\"{$query} 60 IN A") 
+     . " /var/db/pfblockerng/dnsblalias/*", $match);
 $pfb_query = strstr($match[0], 'DNSBL', FALSE);
// Query for a TLD Block
\end{lstlisting}

\section{Threats to Validity}\label{sec:threats}

\noindent\textbf{External Threats.}
A primary external threat pertains to the reproducibility of the baselines. While we endeavored to faithfully implement PatchScout based on its published methodology, the absence of its source code posed challenges. 
To ensure a robust comparison with state-of-the-art methods, we also incorporated BM25 and ColBERT, which are notable sparse and dense retrieval models, respectively, as additional baselines. 


\noindent\textbf{Internal Threats.}
Our dataset's quality and scope could introduce internal threats. To minimize it, we initially built our dataset upon datasets from prior studies~\cite{Tan-ccs21,wang2022vcmatch}, and followed their practice to source CVEs and security patches from various public advisories. Despite our efforts to curate a broad and diverse dataset, biases from these sources might persist. In the future, we will consider manual inspection or automatic techniques that can help assess and improve the data quality.

\section{Related Work}\label{sec:related_work}
There are numerous works focusing on tracing security patches, which can be divided into two categories: tracing security patches for disclosed vulnerabilities~\cite{Tan-ccs21,wang2022vcmatch,Xu-fse22,nguyen2022vulcurator,nguyen2022hermes} and identifying silent security patches~\cite{zhou2017automated,zhou2021spi,zhou2021finding,wu2022enhancing,wang2021patchrnn,cabrera2021commit2vec,icse23-xiaxin}. 

\noindent\textbf{Tracing security patches for disclosed vulnerabilities.} 
Xu et al.~\cite{Xu-fse22} conducted an empirical study to understand the quality and characteristics of patches for disclosed vulnerabilities in two industrial vulnerability databases, thereby proposing to track patches from the CVE reference links across multiple knowledge sources (e.g., Debian). Their work focuses on analyzing reference links provided by security analysts, instead of directly tracing patches from OSS repositories. 
Tan et al.~\cite{Tan-ccs21} conducted the most related work with \ourtool. They designed a ranking-based tool named PatchScout to locate the patch commits by using RankNet on manually defined features from the CVE description and commits. 
Similarly, VCMatch~\cite{wang2022vcmatch}, which directly classifies one commit as related or unrelated to the CVE description by fusing the features from PatchScout and extracted vectors from Bert. 
Unlike PatchScout and VCMatch, 
\ourtool introduces a novel two-phase framework designed to overcome the challenges posed by large search spaces, and enables an end-to-end fine-tuning, to fully exploit the natural correlation between CVE descriptions and commits.

\noindent\textbf{Identifying silent security patches.} 
Besides, several works~\cite{zhou2017automated,zhou2021spi,wu2022enhancing,wang2021patchrnn, zhou2021finding,cabrera2021commit2vec,icse23-xiaxin} have delved into silent security patch identification. These efforts discern security patches but do not correlate them with specific vulnerabilities they rectify. In contrast, our focus is on tracing security patches tailored to a particular vulnerability, as defined by its CVE description.

\section{Conclusion}\label{sec:conclusion}
In this paper, we present \ourtool, an end-to-end and LLM-enhanced two-phase approach for effectively tracing security patches for disclosed vulnerabilities in OSS. The first phase employs a hybrid retriever for the initial retrieval of relevant commits, significantly narrowing down the candidate pool. The second phase leverages a fine-tuned CodeReviewer model to re-rank these commits, achieving a high degree of accuracy. Our extensive evaluations demonstrate that \ourtool consistently outperforms state-of-the-art methods in Recall@K, MRR, and manual efforts, setting \ourtool as a new benchmark in the field of security patch tracing. 



\begin{acks}
This work is supported by the National Key R\&D Program of China under grant 2021ZD0114501, the RIE2020 Industry Alignment Fund – Industry Collaboration Projects (IAF-ICP) Funding Initiative, as well as cash and in-kind contributions from the industry partner(s), the National Research Foundation, Singapore, and the Cyber Security Agency under its National Cybersecurity R\&D Programme (NCRP25-P04-TAICeN). Any opinions, findings and conclusions, or recommendations expressed in this material are those of the author(s) and do not reflect the views of National Research Foundation, Singapore and Cyber Security Agency of Singapore.

\end{acks}

\bibliographystyle{ACM-Reference-Format}
\bibliography{ref}

\end{document}